\documentclass[12pt]{article}
\usepackage{graphicx}

\usepackage{hyperref}    
\usepackage{cite} 
\usepackage{mciteplus}
\usepackage{lineno}

\newcommand\pubnumber{DPF2013-87}
\newcommand\pubdate{\today}

\def\napoli{CPPM, Aix-Marseille Universit\'e CNRS/IN2P3, Marseille, France}

\def\Title#1{\begin{center} {\Large #1 } \end{center}}
\def\Author#1{\begin{center}{ \sc #1} \end{center}}
\def\Address#1{\begin{center}{ \it #1} \end{center}}

\newcommand\pubblock{\rightline{\begin{tabular}{l} \pubnumber\\
         \pubdate  \end{tabular}}}
\newenvironment{Abstract}{\begin{quotation}  }{\end{quotation}}
\newenvironment{Presented}{\begin{quotation} \begin{center} 
             PRESENTED AT\end{center}\bigskip 
      \begin{center}\begin{large}}{\end{large}\end{center} \end{quotation}}
\def\Acknowledgments{\bigskip  \bigskip \begin{center} \begin{large}
             \bf ACKNOWLEDGMENTS \end{large}\end{center}}

\def\beq{\begin{equation}}
\def\eeq#1{\label{#1}\end{equation}}
\def\eeqn{\end{equation}}

\def\beqa{\begin{eqnarray}}
\def\eeqa#1{\label{#1}\end{eqnarray}}
\def\eeqan{\end{eqnarray}}

\let\bar=\overbar

\def\ket#1{\left| {#1} \right\rangle}

\def\Dslash{\not{\hbox{\kern-4pt $D$}}}
\def\dslash{\not{\hbox{\kern-2pt $\del$}}}

\def\msb{{\bar{\ssstyle M \kern -1pt S}}}

\newcommand{\comment}[1]{\par\noindent {\em\small [#1]}}
\renewcommand{\comment}[1]{}

\newcommand{\unit}[1]{\ensuremath{\rm\,#1}}

\newcommand{\mevcc}{\unit{MeV\!/\!{\it c}^2}}

\newcommand{\TeV}{\unit{TeV}}
\newcommand{\invfb}{\unit{fb^{-1}}}

\newcommand{\invps}{\unit{ps^{-1}}}

\newcommand{\fs}{\unit{fs}}

\newcommand{\rad}{\unit{rad}}

\newcommand{\CP}{\ensuremath{\rm CP}}

\newcommand{\DGs}{\ensuremath{\Delta\Gamma_{ s}}}

\newcommand{\Gs}{\ensuremath{\Gamma_{ s}}}

\newcommand{\phis}{\ensuremath{\phi_{ s}}}
\newcommand{\betas}{\ensuremath{\beta_{ s}}}

\newcommand{\Vcs}{\ensuremath{V_{ cs}}}
\newcommand{\Vcb}{\ensuremath{V_{ cb}}}

\newcommand{\Vts}{\ensuremath{V_{ ts}}}
\newcommand{\Vtb}{\ensuremath{V_{ tb}}}

\newcommand{\etag}{{\ensuremath{\varepsilon_{\rm tag}}}}

\newcommand{\effD}{{\ensuremath{\etag D^2}}}

\newcommand{\BAR}[1]{\overline{#1}}

\newcommand{\particle}[1]{{\ensuremath{#1}}}

\newcommand{\Bs}{\particle{B^0_s}}

\newcommand{\Bsbar}{\particle{\BAR{B}{^0_s}}}

\newcommand{\jpsi}{\particle{J\!/\!\psi}}
\newcommand{\Jpsi}{\particle{J\!/\!\psi}}

\newcommand{\Kplus}{\particle{K^+}}
\newcommand{\Kp}{\particle{K^+}}
\newcommand{\Kminus}{\particle{K^-}}
\newcommand{\Km}{\particle{K^-}}

\newcommand{\pip}{\particle{\pi^+}}
\newcommand{\pim}{\particle{\pi^-}}

\newcommand{\Jhh}{\particle{J\!/\!\psi h h}}
\newcommand{\BsBs}{\Bs--\Bsbar}

\newcommand{\decay}[2]{\particle{#1\!\to #2}}

\newcommand{\BsJKK}{\decay{\Bs}{\Jpsi\Kplus\Kminus}}             
\newcommand{\BsJpp}{\decay{\Bs}{\Jpsi\pip\pim}}             
           
\newcommand{\delperp}{\delta_{\perp}}
\newcommand{\delpar}{\delta_{\|}}

\def\apar{A_{\|}(t)}
\def\aperp{A_{\perp}(t)}
\def\azero{A_{0}(t)}

\newcommand{\aperpsq}{\ensuremath{|\aperp|^2}}
\newcommand{\aparsq}{\ensuremath{|\apar|^2}}
\newcommand{\azerosq}{\ensuremath{|\azero|^2}}

\def\maglambda{|\lambda|}

\begin{document}
\begin{titlepage}
\pubblock

\vfill
\Title{Measurement of \phis\ at LHCb}
\vfill
\Author{Olivier Leroy}
\begin{center}
{\small on behalf of the LHCb Collaboration}
\end{center}
\Address{\napoli}
\vfill
\begin{Abstract}
The study of CP violation in \Bs\ oscillations is one of the key goals of the LHCb experiment. Effects are predicted to be very small in the Standard Model but can be significantly enhanced in many models of new physics. We present the world's best measurement of the CP-violating phase \phis\ using \BsJKK\ and \BsJpp\ decays.

\end{Abstract}
\vfill
\begin{Presented}
DPF 2013\\
The Meeting of the American Physical Society\\
Division of Particles and Fields\\
Santa Cruz, California, August 13--17, 2013\\
\end{Presented}
\vfill
\end{titlepage}
\def\thefootnote{\fnsymbol{footnote}}
\setcounter{footnote}{0}
%

\section{Introduction}

The interference between \Bs-mesons decaying to a \Jhh\ final-state either directly of via \BsBs\ oscillation gives rise to a CP-violating phase called \phis\ $(h=K~ {\rm or}~ \pi)$. In the Standard Model, neglecting sub-leading penguin contributions, this phase is predicted to be $-2\betas$, where $\betas =\arg\left(-\Vts\Vtb^*/\Vcs\Vcb^*\right)$ and $V_{ij}$ are elements of the CKM matrix\cite{CKM}. 
 The indirect determination via global fits to experimental data gives
 $2\betas=0.0364\pm0.0016\rad$~\cite{CKMfitter}. This can be significantly modified if New Physics contributes to the \BsBs\ mixing box. 
Direct measurements have been reported by LHCb~\cite{LHCb:2011aa, phisJpipi}, ATLAS~\cite{atlas:2012fu}, CDF~\cite{Aaltonen:2012ie} and D0~\cite{Abazov:2011ry}. 
This document summarizes the \BsJKK\ and \BsJpp\ analyses performed by LHCb, using 1\invfb\ of data collected in 2011 at a center-of-mass energy of 7\TeV~\cite{LHCbPaper2013-002}. 

\section{\BsJKK\ analysis}

The weak phase \phis\ is extracted using a tagged-time-dependent angular fit to \BsJKK\ candidates. 
 The final state is decomposed into four amplitudes: three P-wave, $A_0, A_{\parallel} , A_\perp$ and one S-wave, $A_S$ accounting for the non-resonant $K^+K^-$ configuration. It is a superposition of  CP-even states, $\eta_i =  +1$ for $i\in\{0,\parallel\}$ and CP-odd states, $\eta_i =-1$ for $i\in\{\perp,{\rm S}\}$ states. 
The amplitudes are parameterised by $|A_i|e^{i\delta_{i}}$ with the conventions $\delta_0=0$ and $\azerosq + \aparsq + \aperpsq  = 1$.
The phase $\phis$ is  defined by  $\phis = -\arg( \lambda )$, where  $\lambda= \lambda_i/\eta_i$ and  $\lambda_i \; = \; \frac{q}{p} \; \frac{\bar{A}_i}{A_i}$. 
The complex parameters $p$ and $q$ describe the
relation between mass and flavour eigenstates: $\ket{B_{L,H}} = p \ket{\Bs} \pm q \ket{\Bsbar}$ and  $p^2+q^2=1$.

The reconstruction of \BsJKK\ candidates proceeds using the decays
\mbox{$\Jpsi\to\mu^{+}\mu^{-}$} combined with a pair of oppositely charged kaons.
The invariant mass of the $\Kp\Km$ pair must be in the
range $[990, 1050]\mevcc$.
 After the trigger and full offline selection, we find
$27\,617 \pm 115$ $\Bs\to\Jpsi K^{+}K^{-}$ \BsJKK\  signal candidates.
The decay time resolution is estimated event by event, with scale factor determined on real data using prompt \Jpsi$K^+K^-$ combinations. On average, it is 45\fs.
The decay time acceptance is determined on real data, using a prescaled unbiased trigger sample. 
The angular acceptance is determined using Monte Carlo sample. 
The flavour tagging algorithm uses information from the same-side and opposite-side with respect to the signal candidate. It is optimised on Monte Carlo samples and calibrated on real data, using flavor specific control channels. 
The overall effective tagging power obtained is \mbox{$\effD = (3.13 \pm 0.12 \pm 0.20)$}\%, the tagging
efficiency is $\etag=(39.36\pm0.32)$\% and the wrong-tag probability is $\omega = 35.9$\%.

\begin{figure}[htb]
  \begin{center}
    \includegraphics[width=0.42\textwidth]{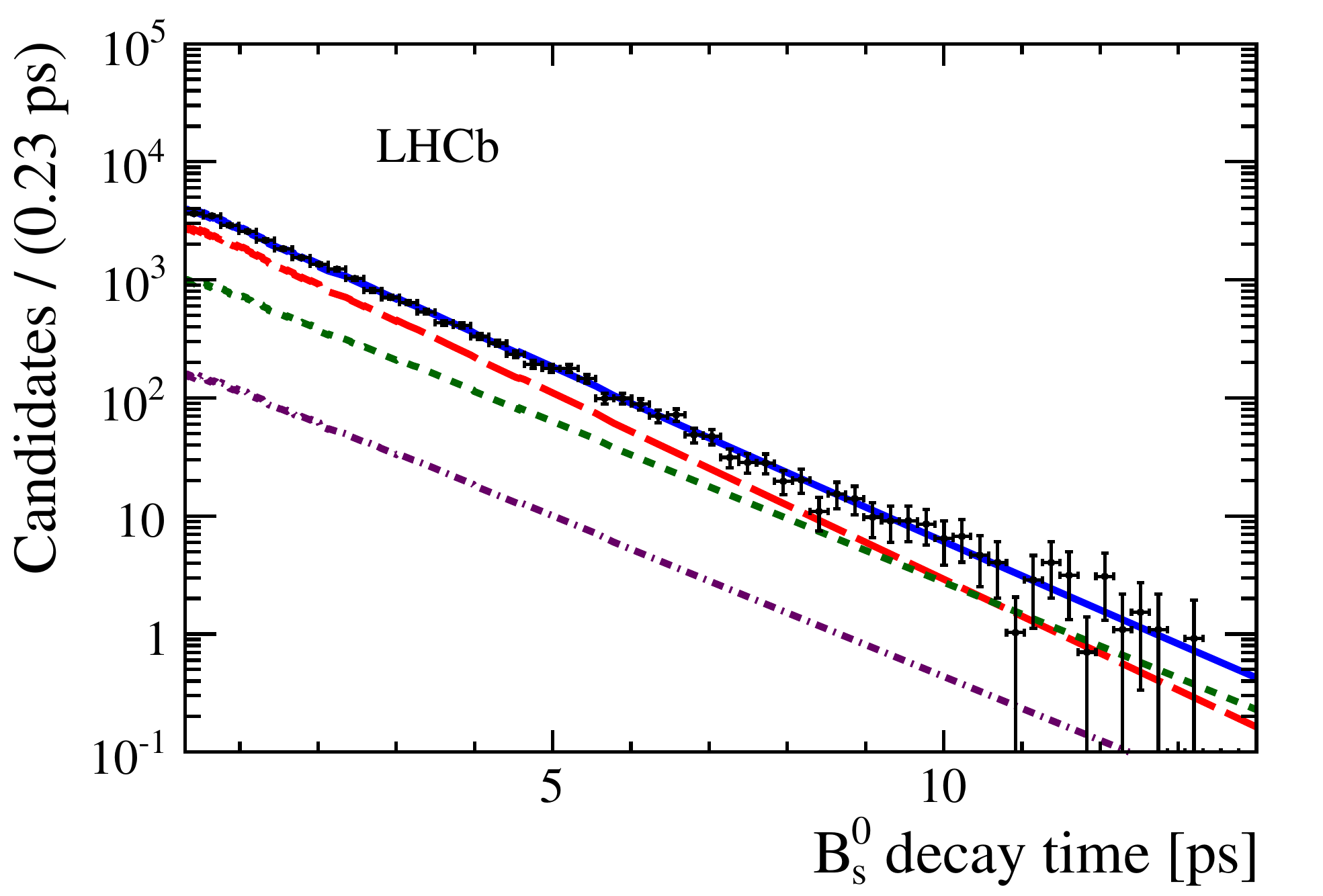}
    \includegraphics[width=0.42\textwidth]{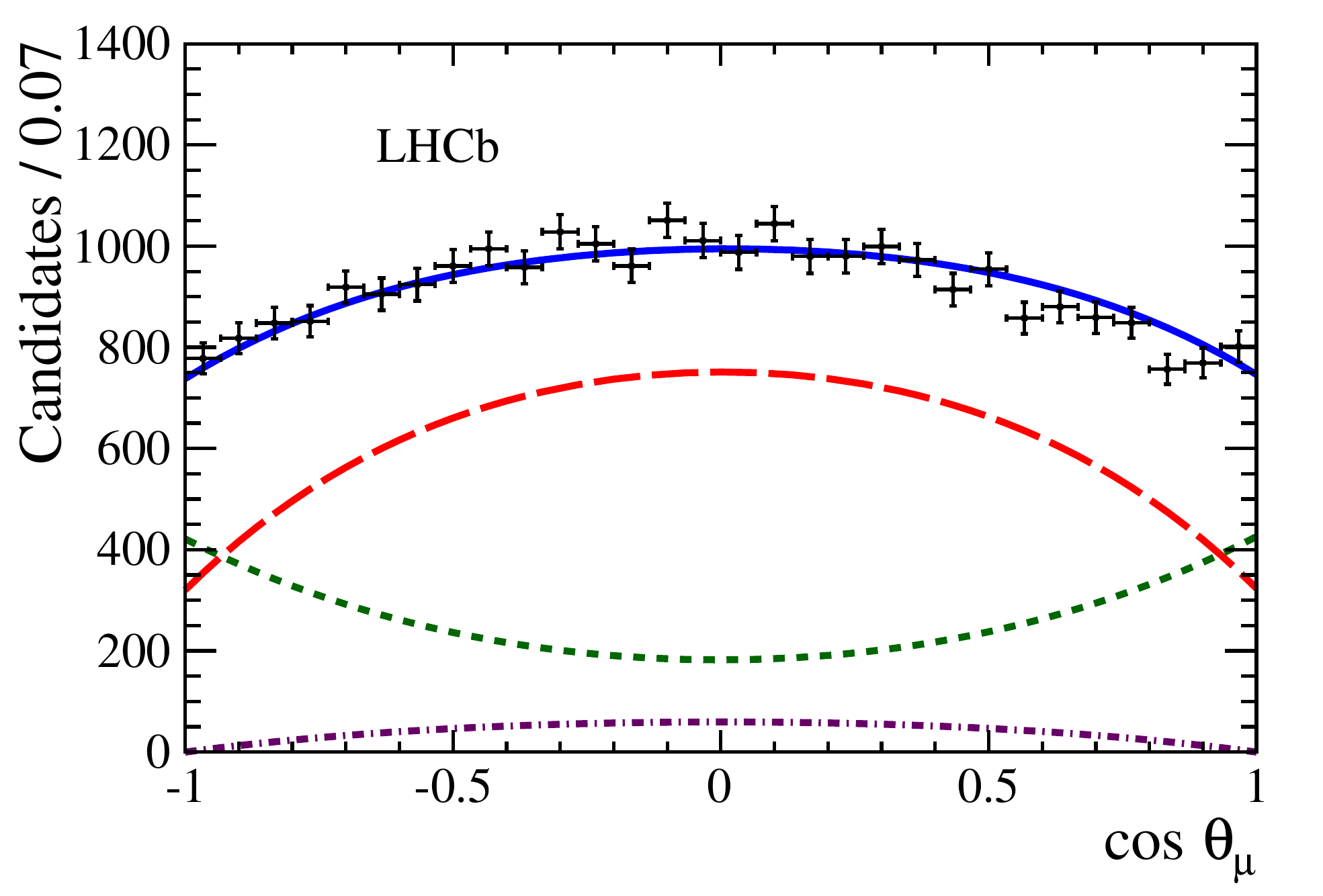} \\
    \includegraphics[width=0.42\textwidth]{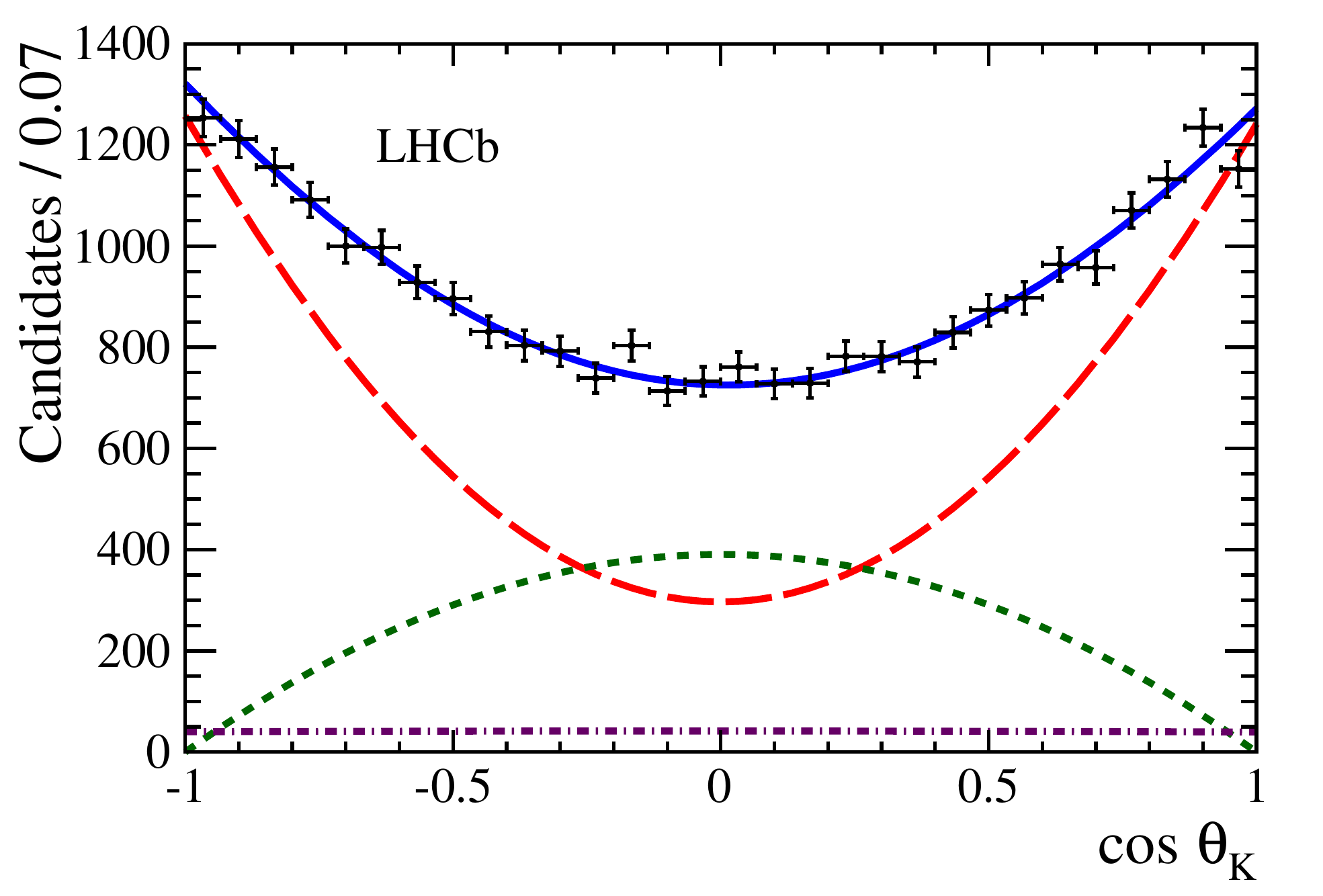}
    \includegraphics[width=0.42\textwidth]{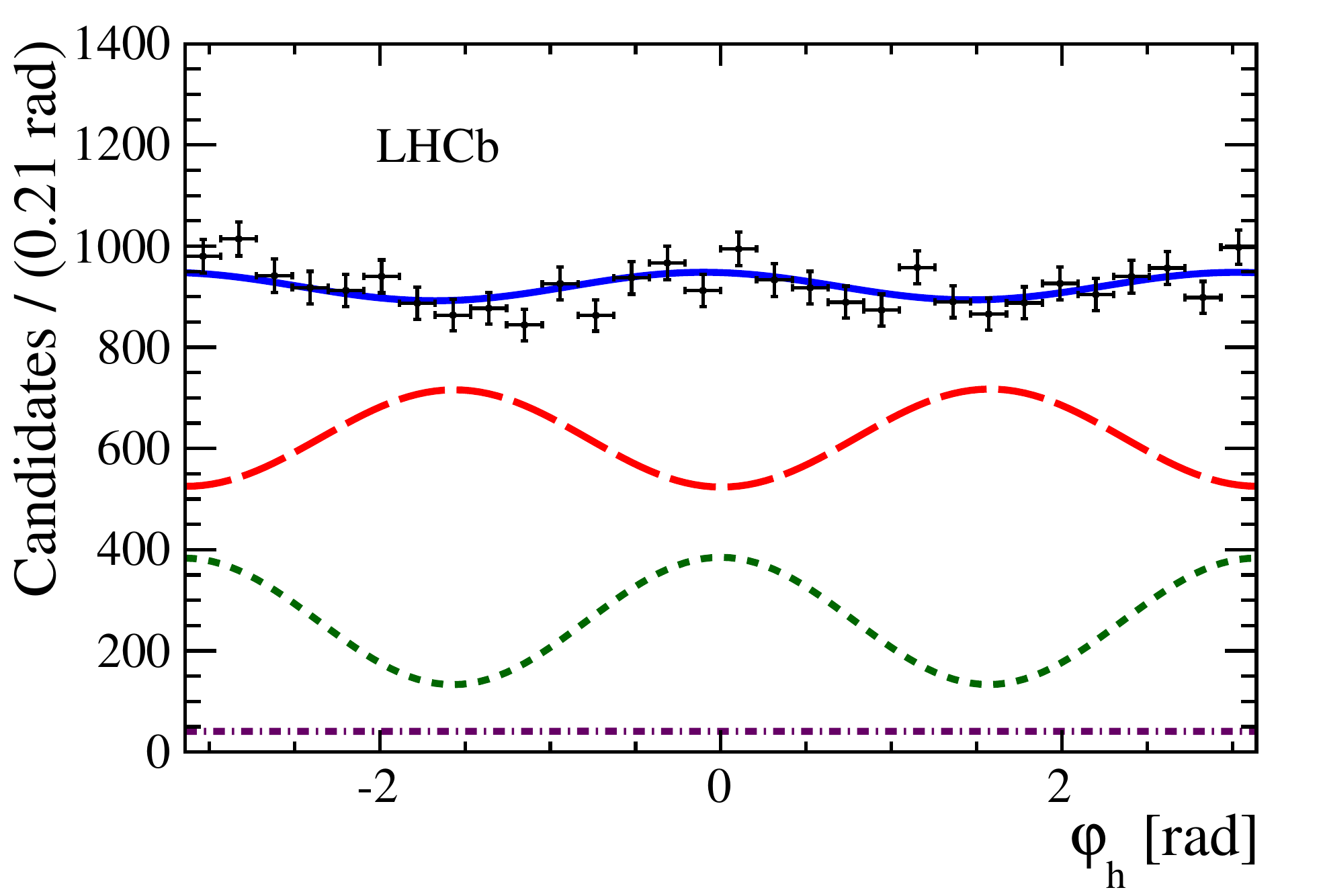}
  \end{center}
%
%
\caption{\small\label{fig:results-projections} 
Decay-time and helicity-angle distributions for $\Bs\to\jpsi\Kp\Km$ decays (data points) with the one-dimensional projections of the PDF at the maximal likelihood point. 
The solid blue line shows the total signal contribution, which is composed of 
\CP-even (long-dashed red), \CP-odd (short-dashed green) and S-wave (dotted-dashed purple) contributions.}
\label{fig:proj}
\end{figure}

A weighted unbinned likelihood fit is performed using a signal-only PDF, as described in~\cite{sfit}. 
The signal weights are determined using the $sPlot$ method~\cite{sPlot}. 
The data is divided into six independent invariant $K^+ K^-$ mass bins. This improves the statistical sensitivity and allows to resolve the two-fold ambiguity of \BsJKK\ differential decay rate, in particular the sign of \DGs, as in~\cite{dgspaper}.
The projections of the decay time and angular distributions are shown in Fig.~\ref{fig:proj}. 
The principal results of the maximum likelihood fit are:
 $\phi_s = 0.07  \pm 0.09  \rm{(stat)} \pm 0.01 \rm{(syst)}\ \rm{rad}$,
  $\Gamma_s = 0.663  \pm  0.005  \rm{(stat)} \pm 0.006 \rm{(syst)}\ \invps$ and
  $\DGs  = 0.100   \pm  0.016    \rm{(stat)} \pm  0.003  \rm{(syst)}\ \invps$,
The other results of the fit are presented in the next section. 

\section{\BsJpp\  analysis and combined results}

The \BsJpp\ analysis is similar to the \BsJKK\ one with a noticeable simplification: the final state being CP-odd, there is no need for the angular analysis. 
After trigger and selection, 2936 \BsJpp\ signal candidates are used in the analysis. 
The decay time resolution is 40\fs\ and the effective tagging power is 3.37\%. 
The results of the simultaneous fit to both \BsJKK\ and \BsJpp\ are given in Table~\ref{tab-app-combined-results-II}.

\begin{table}[htb]
\begin{center}\small
    \begin{tabular}{l|c}
      Parameter            &  Value\\
      \hline
      $\Gs$ [\invps]   	& $0.661		\pm 0.004 			\pm  0.006$\\ 
      $\DGs$ [\invps]  	& $0.106		\pm 0.011			\pm 0.007$\\
      $\aperpsq$        	& $0.246 		\pm 0.007 			\pm  0.006$\\
      $\azerosq$           & $0.523 		\pm 0.005 			\pm 0.010$\\
      $\delpar$ [rad] 	& $3.32 	\,	^{+0.13}_{-0.21} \pm  0.08$\\
      $\delperp$ [rad]   & $3.04 		\pm 0.20 			\pm 0.08$\\ 
      $\phis$ [rad]         & $0.01 		\pm 0.07 			\pm 0.01 $\\
     $\maglambda$	& $0.93		\pm0.03			\pm0.02 $\\
    \end{tabular}
    \caption{\small Results of combined fit to the \BsJKK\ and \BsJpp\ datasets. 
      The first uncertainty is statistical and the second is systematic.}  
    \label{tab-app-combined-results-II}
  \end{center}
\end{table}

\begin{figure}[htb]
 \begin{center}
    \includegraphics[width=0.9\textwidth]{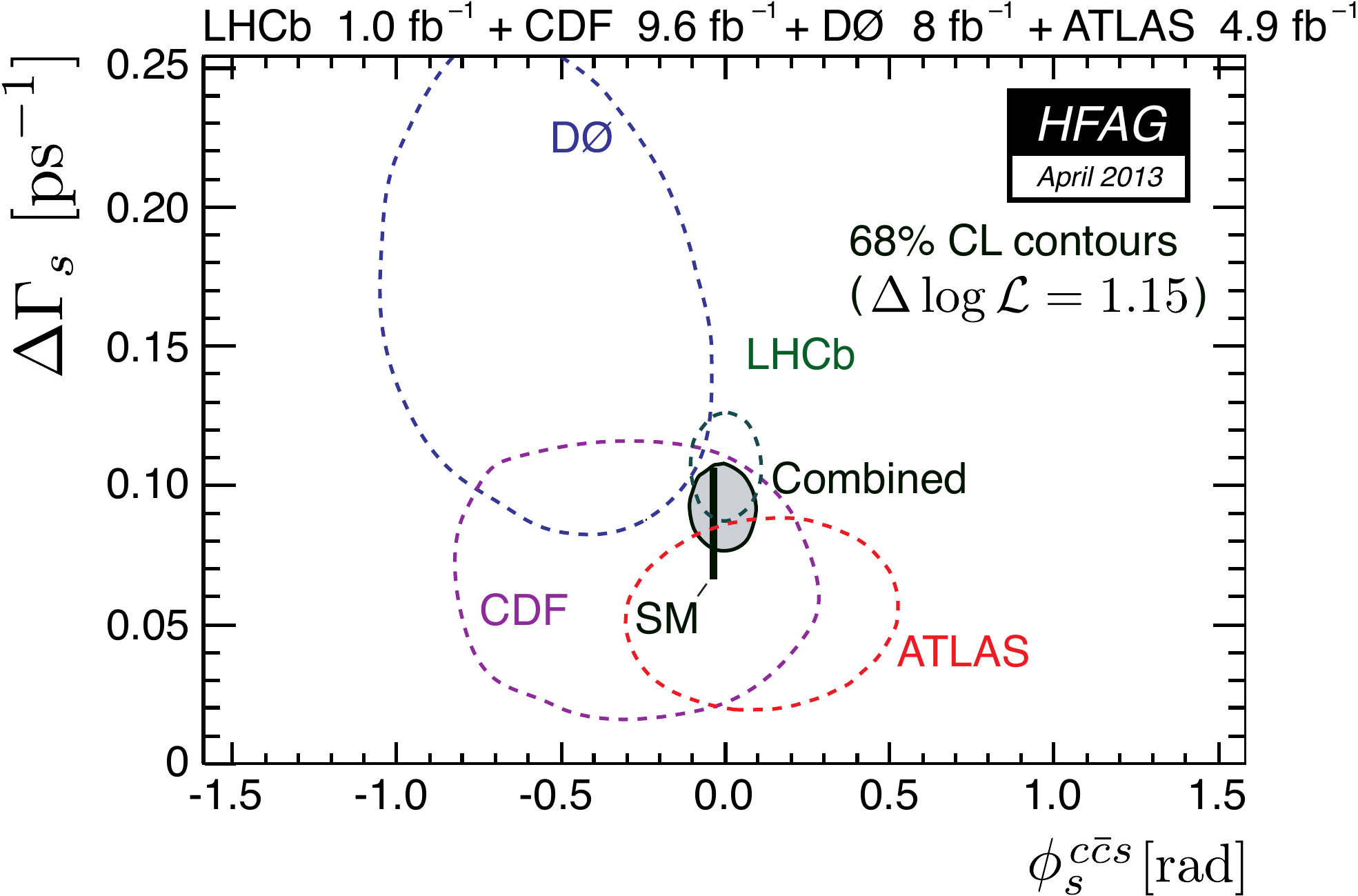}
  \end{center}
\caption{\small 68\% CL regions in \DGs\ and \phis\ obtained from individual and combined CDF, D0, LHCb and ATLAS likelihoods of \BsJKK\ and \BsJpp\ samples~\cite{HFAG}. The expectation from the Standard Model~\cite{CKMfitter, Lenz} is shown as a black triangle.} 
\label{fig:hfag}
\end{figure}

The total systematics uncertainty on \phis\ is 0.01, seven times smaller than the statistical one. It is dominated by the size of the Monte Carlo sample used to determine the angular acceptance, hence it will even reduce in the future. The systematics on \DGs\ are also small. They are dominated by the background subtraction method and the decay time acceptance. The only fitted parameter which uncertainty is dominated by systematical effect is \Gs. This is due to the decay time acceptance determination and it will also decrease in the near future. 

A comparison of the LHCb result with those of other experiments is shown in Fig.~\ref{fig:hfag}. 

\section{Conclusions}
    With 1\invfb, using \BsJKK\ and \BsJpp,
    LHCb performed the world most precise measurements of:
      \[
      \setlength{\arraycolsep}{0.5mm}
      \begin{array}{ccllllllll}
        \phi_s &\;=\; & 0.01  &\pm & 0.07  & \rm{(stat)} &\pm & 0.01 & \rm{(syst)} & \rm{rad},\\
        \rule{12.9mm}{0mm}\Gamma_s \rule{12.9mm}{0mm}  &\;=\; & 0.661  &\pm & 0.004 & \rm{(stat)} &\pm & 0.006 & \rm{(syst)} & \invps \rule{0pt}{5mm} \\
        \DGs  &\;=\; & 0.106   &\pm & 0.011    & \rm{(stat)} &\pm & 0.007 & \rm{(syst)} & \invps \rule{0pt}{5mm} \\
      \end{array}
      \]

    The results are compatible with Standard Model so far and put stronger constraints than ever on possible Standard Model  extensions in the \BsBs\ mixing phase. Though, there is still room for new physics. 
The uncertainty on \phis\ is expected to decrease to $\sim0.05$ by the end of 2013, thanks to the 3\invfb\ of data already registered by LHCb, and to $\sim0.008$ by the end of the LHCb upgrade phase.

\newpage
\Acknowledgments
I would like to thanks the organizers of the DPF'2013 for the nice atmosphere 
during the conference in Santa Cruz and my LHCb colleagues who helped in the preparation of this talk.



\begin{thebibliography}{99}

\bibitem{CKM} T. Maskawa and M. Kobayashi, Prog. Th. Phys {\bf 49} (1973) 652. N. Cabibbo, Phys. Rev. Lett. {\bf 10} (1963) 531. 

\bibitem{CKMfitter} J. Charles et al., CKMfitter, Phys. Rev. D84 033005 (2011), with
 updated results and plots available at \url{http://ckmfitter.in2p3.fr}.

\bibitem{LHCb:2011aa}
R. Aaij {\it et al.}, LHCb Collaboration, Phys. Rev. Lett. {\bf 108} (2012) 101803.

\bibitem{phisJpipi}
R. Aaij {\it et al.}, LHCb Collaboration, Phys. Lett. {\bf B713} (2012) 378-386. 

\bibitem{atlas:2012fu}
G. Aad {\it et al.}, ATLAS Collaboration, JHEP 12 (2012) 072. 

\bibitem{Aaltonen:2012ie}
T. Aaltonen {\it et al.}, CDF Collaboration,  Phys. Rev. Lett. {\bf 109} (2012) 171802. 

\bibitem{Abazov:2011ry}
V.M. Abazov   {\it et al.},   D0  Collaboration,  Phys. Rev. {\bf D85} (2012) 032006.


\bibitem{LHCbPaper2013-002}
R. Aaij {\it et al.}, LHCb Collaboration, Phys. Rev. {\bf D87} (2013) 112010. 

\bibitem{sfit} Y. Xie, \href{http://arxiv.org/abs/0905.0724}{arXiv:0905.0724} (2009).

\bibitem{sPlot} M. Pivk and F. Le Diberder, Nucl. Instrum. Math. {\bf A555} (2005) 356. 

\bibitem{dgspaper} R. Aaij {\it et al.}, LHCb Collaboration,  Phys. Rev. Lett. {\bf 108} (2012) 241801. 

\bibitem{HFAG} Y. Amhis {\it al.},
\href{http://arxiv.org/abs/1207.1158}{arXiv:1207.115} (2012)
and online update at \url{http://www.slac.stanford.edu/xorg/hfag}.

\bibitem{Lenz}
A. Lenz, U. Nierste, \href{http://arxiv.org/abs/1102.4274}{arXiv:1102.4274} (2011). 


\end{thebibliography}
\end{document}